\documentclass{aa}

\usepackage{natbib}
\usepackage{graphicx}
\usepackage{txfonts}

\begin{document}

\title{Magnetohydrostatic (MHS) atmospheres}

\author{Laurence J. November}

\institute{The Light Physics, La Luz NM 88337-0217 USA}

\titlerunning{MHS atmospheres}
\authorrunning{L. J. November}
\mail{laluzphys@yahoo.com}
\date{December 24, 2003} 

\abstract{
 We show that the atmospheric and magnetic height variations are
coupled in general MHS equilibria with gravity when isolated thin
non-force-free flux tubes are present.  In gas-dominated
environments, as in stellar photospheres, flux tubes must expand
rapidly with height to maintain pressure balance with the cool
surroundings. But in magnetically dominated environments, as in
stellar coronae, the large-scale background magnetic field determines
the average spreading of embedded flux tubes, and rigidly held flux
tubes {\it require} a specific surrounding atmosphere with a unique
temperature profile for equilibrium.  The solar static equilibrium
atmosphere exhibits correct transition-region properties and the
accepted base coronal temperature for the sun's main magnetic
spherical harmonic.  Steady flows contribute to the overall pressure,
so equilibria with accelerated wind outflows are possible as well.
Flux tubes reflect a mathematical degeneracy in the form of
non-force-free fields, which leads to coupling in general equilibrium
conditions.  The equilibrium state characterizes the system average
in usual circumstances and dynamics tend to maintain the MHS
atmosphere. Outflows are produced everywhere external to rigidly held
flux tubes that refill a depleted or cool atmosphere to the
equilibrium gas profile, heating the gas compressively.
 }

\maketitle

\section{Introduction}
\label{s:intro}

\begin{figure}
\centering
\includegraphics[width=8.5cm]{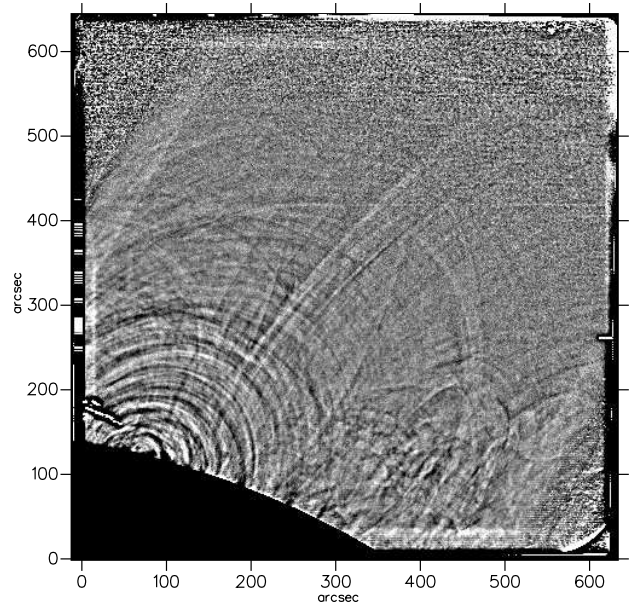}
 \caption{Unsharp-masked time-averaged white-light eclipse image
processed by subtracting a 6 arcsec = 4300 km Gaussian convolution
shown with contrast range $\Delta \ln I = 0.8\%$.  Dark and bright
threads are probably flux tubes in current sheets.}
 \label{f:cfht}
\end{figure}

In our observation of solar `threads'
(\citealt{November+Koutchmy1996}, hereafter NK96) we puzzled over
whether coronal flux tubes should expand with height to maintain
pressure balance with the ambient gas-pressure decrease or follow the
lines of the external magnetic field.  After all, if the corona is
independently heated, the gas and magnetic height variations should
be generally different.  The many observations of EUV and X-ray loops
as well as the purely density-sensitive observations of white-light
$>15$ Mm `voids' (\citealt{MacQueen+++1974a}), $>2$ Mm threads, and
scintillation $>1$ {\it kilometer} filamentary microstructure
(\citealt{Coles+Harmon1978, WooR+++1995}) all suggest that the
large-scale solar coronal magnetic field is interspersed with flux
tubes with a great range of sizes.  Figure \ref{f:cfht} shows dark
and bright threads in a square region about a solar radius on a side
over the west limb taken with the Canada-France-Hawaii Telescope
(CFHT) at the unique total-eclipse opportunity on Mauna Kea on July
11, 1991.  The dark and bright threads, which were essentially
unchanged over the 4-minute eclipse duration, appear to be aligned
and organized in arched and radial surfaces, probably current sheets,
overlapping in their projection onto the plane of the sky in the
line-of-sight view.

We know that isolated thin flux tubes must expand with height in
balance with the ambient gas-pressure decrease.  The flux-tube gas
pressure difference, external minus internal, is related to the
magnetic pressure difference by the static equilibrium relation
 \noindent
 \begin{equation}
p_e-p_i={1\over8\pi}\left(\vec{B}_i^2-\vec{B}_e^2\right).
\label{e:pdif}
\end{equation}
 The Wilson depression in sunspots shows that at least large-scale
photospheric flux tubes are nearly evacuated.  Evacuated flux tubes
exhibit the maximum magnetic flux concentration and field strength
possible.  In evacuated flux tubes $p_i=0$ and the photospheric
external field $\vec{B}_e=0$, so the internal field strength must
follow the ambient gas-pressure decrease $\vec{B}_i(r)^2= 8\pi
p_e(r)$, giving flux tubes that expand with height to conserve their
total internal magnetic flux.  Since the scale height for $p_e(r)$ is
small in the relatively cool photosphere and chromosphere, a dramatic
flux-tube expansion occurs, which is the accepted explanation for the
solar `magnetic canopy' (\citealt{Giovanelli1982}).  Both voids and a
range of sizes of dark threads were also found to have contrasts
consistent with what is expected for fully evacuated cylindrical flux
tubes (\citealt{MacQueen+Sime+Picat1983}; NK96).  In the corona,
where the external magnetic field is not zero, the magnetic flux
inside evacuated flux tubes must decrease with height to follow the
total external ambient gas plus magnetic pressures, or
$\vec{B}_i(r)^2= 8\pi p_e(r)+\vec{B}_e(r)^2$.  This gives a small
flux-tube expansion with height in the large gas-pressure scale
height of the hot corona.  A slight flux-tube expansion with height
seems evident in some of the threads in Fig. \ref{f:cfht}.

We can argue too that the radial expansion of isolated thin flux
tubes must conform to the external magnetic field.
\cite{ParkerEN1972} shows that
 \begin{equation}
\left(\vec{B}_e\vec{\cdot}\vec{\nabla}\right) \vec{B}_i=0,
\label{e:P72}
\end{equation}
 in MHS equilibria without gravity, where $\vec{B}_e$ represents a
uniform background magnetic field, and the field in flux tubes
$\vec{B}_i$ contains all of the spatial variations.  The theorem is
locally applicable everywhere in perturbed MHS equilibria with
gravity, i.e. Eq. (\ref{e:P72}) with $\vec{B}_e$ allowed to vary
slowly spatially, as we discuss in Sect. \ref{s:MHSg}.  Thus thin
flux tubes in a gravitational field must spread to follow the lines
of the relatively uniform background magnetic field; the radial
functional variations must be directly related, that is
$|\vec{B}_i(r)| \propto |\vec{B}_e(r)|$ along flux tubes.

Therefore both ways of looking at the problem are essentially
correct.  The coronal flux-tube field must comply with the external
field $\vec{B}_i(r)^2 \propto \vec{B}_e(r)^2$, and the radial
variation of the external gas pressure outside isolated thin
evacuated coronal flux tubes is proportional to the magnetic pressure
difference $p_e(r) \propto \vec{B}_i(r)^2- \vec{B}_e(r)^2$. Hence,
the external gas and magnetic pressures must be coupled with $p_e(r)
\propto \vec{B}_e(r)^2$ outside and along isolated thin evacuated
flux tubes.  The background magnetic field is dominant in lower
stellar coronae and must determine the expansion of embedded flux
tubes.  The expansion of evacuated flux tubes then defines the radial
variation of the ambient gas pressure for equilibrium, the gas
pressure being horizontally stratified in the external force-free or
potential field.  More generally, flux tubes of arbitrary base
pressure but in thermal balance with the local ambient atmosphere
satisfy $p_i(r)\propto p_e(r)$, giving the same external atmosphere
$p_e(r) \propto \vec{B}_e(r)^2$. Thus an atmosphere containing
isolated thin non-force-free flux tubes in different forms of local
balance must exhibit a temperature profile defined by the dominant
background magnetic field for a perfect gas in hydrostatic
equilibrium.

The properties of flux tubes reflect a natural degeneracy in the form
of non-force-free magnetic fields, and imply the existence of an
underlying {\it mathematical} relationship between an equilibrium
atmosphere and large-scale magnetic field.  In Sect. \ref{s:MHSg}, we
show that the height variations of the gas pressure and magnetic
field amplitude are directly coupled in the most-significant-order
terms of the gravitational perturbation in general non-force-free MHS
equilibria, with smaller field divergence and curl effects.  In
low-$\beta$ systems, with strong magnetic fields, the external gas
pressure corresponds to a small difference $p_e(r)= (\vec{B}_i(r)^2-
\vec{B}_e(r)^2)/(8\pi)$, which allows large pressure deviations for a
given relative flexibility of the large-scale magnetic field around
individual flux tubes.  Magnetic fields can exhibit and persist in
significant deviations from equilibrium, however the coupling $p_e(r)
\propto \vec{B}_e(r)^2$ describes the general equilibrium condition
for arbitrary $\beta$.

Historical MHS equilibrium solutions with gravity have also suggested
that a general relationship exists between the large magnetic and
gas-pressure scales.  \cite{Dungey1953a} derived plane-parallel MHS
equilibria for a magnetic field with its degenerate axis directed
perpendicular to gravity (see also: \citealt{LowBC1975a,
ChengCZ+ChoeGS1998}).  His solutions are represented in a modified
Grad-Shafranov (GS) Equation that retains the gas-pressure radial
scale-height variation as a separable factor in the pressure term. 
Dungey showed that when the magnetic field is also zero in its
degenerate axis, the separable factor can be absorbed into the
coordinates, giving the classical GS Equation in a remapped
coordinate system. The new GS coordinates are defined by the
gas-pressure scale height, meaning that magnetic fields in a
prespecified atmosphere must be spatially distorted.  Horizontal flux
tubes exhibit substantial radial deformation in a solar corona of
given constant temperature (\citealt{Zweibel+Hundhausen1982,
LowBC1992a}). Depending upon their size compared to the gas-pressure
scale height, flux tubes are either compressed or expanded in the
gravitational direction, and approximately undistorted only in a
specific atmosphere.

In Sect. \ref{s:atm}, we examine equilibrium radial profiles of
temperature and flux-tube diameter using an approximate solution to
the equilibrium equations.  The profiles reproduce the salient
features of the solar atmosphere: the rapid expansion of flux tubes
in the photosphere where the atmospheric gas pressure is dominant, a
transition region of reasonable thickness located just above the
height where the gas and magnetic energy densities are equal, and a
corona of correct base temperature for the sun's main magnetic
spherical harmonic.  Steady flows do not alter the coupling, but may
produce a Bernoulli pressure, which modifies the gas pressure, and
leads to possible accelerated steady wind solutions.

Although the dynamics of the system can be quite complicated, the
equilibrium state characterizes the average in many types of MHD
systems, e.g. `quasi-steady' (\citealt{LowBC1980a}). Hydrodynamics in
such systems tend to maintain the MHS atmosphere, as we discuss in
Sect. \ref{s:MHD}.  The possible importance of MHS atmospheres for
other problems in physics and astrophysics beyond the formation of
solar and stellar coronae and in more marginal non-quasi-steady
physical conditions needs to be considered, but discussion lies
beyond the scope of this paper.

\section{MHS equilibria with gravity}
\label{s:MHSg}

The equations for static equilibrium in a gravitational field are the
MHS equation and Gauss' Law
 \begin{eqnarray}
&&\hskip -.3cm\vec{\nabla} p+{p\over h}\vec{\nabla} r={1\over4\pi}(\vec{\nabla}\vec{\times}\vec{B})\vec{\times}\vec{B},
\label{e:MHSg}\\
&&\hskip -.3cm\vec{\nabla}\vec{\cdot} \vec{B}=0.
\label{e:divB}
\end{eqnarray}
 The static contribution to the pressure scale height $h$ is directly
related to the temperature for a perfect gas, $h\equiv p/(\rho
g(r))=kT/ (\bar{m}g(r))$.  The variables have their usual meanings:
$\vec{B}$ is the magnetic field vector in Gaussian units, $p$ the gas
pressure, $\rho$ the density, $T$ the temperature, $\bar{m}$ the mean
particle mass, $k$ Boltzman's constant, $c$ the speed of light (used
below); $g(r)= GM_s/ r^2$ is the gravitational acceleration, with $G$
the universal gravitational constant and $M_s$ the stellar mass. The
coordinate $r$ measures the distance from the system or stellar
center, and the formal notation $\vec{\nabla} r$ is adopted for
algebraic convenience to denote the radially directed unit vector.

The set of four vector-element Eqs. (\ref{e:MHSg}) and (\ref{e:divB})
contain five unknowns, $p$, $h$, and the three vector elements of
$\vec{B}$.  However, the solution space for $\vec{B}$ is more
restrictive than would be obtained with linear relations, as we
discuss in this Section and elaborate in Appendix \ref{a:MHS}.

The gravitational term in Eq. (\ref{e:MHSg}) compared to the pressure
gradient is of order $d/h$, where $d$ is a characteristic flux-tube
thickness; for the main power in threads in the solar corona $d/h <
10^{-3}$.  In the vicinity of non-force-free fields gas-pressure
changes must be mainly magnetically determined, and only far away can
the gravitational gradient be significant.  Where the Lorentz term is
negligible, the gas pressure is hydrostatic and defined by a single
radial temperature profile.  Since the gravitational term is small in
non-force-free conditions, the classical equilibrium solutions
without gravity described in Appendices \ref{a:MHS} and
\ref{a:MHSprop} are applicable, and Eqs. (\ref{e:pdif}) and
(\ref{e:P72}) are valid in every small volume, but with the allowance
that quantities may vary slowly spatially.  Thus we obtain coupled
atmospheric and magnetic height variations, at least if certain types
of flux tubes are present as described in the Introduction.

Conditions along field lines are represented by a parallel-field
equation
 \begin{equation}
\vec{B}\vec{\cdot}\left(\vec{\nabla} p+{p\over h}\vec{\nabla} r\right)=0,
\label{e:B.MHSg}
\end{equation}
 which is the projected component of the MHS Eq. (\ref{e:MHSg}) along
$\vec{B}$.  The relation says that gas pressure changes along field
lines must be hydrostatic, which can be seen by introducing the
integrating factor $\phi$ into the gas pressure
 \begin{equation}
p=\phi \hat{p},
\label{e:pdef}
\end{equation}
 for an unrestricted function $\hat{p}$.  Upon substitution
 \begin{equation}
\vec{B}\vec{\cdot}\left(\phi\vec{\nabla}\hat{p}+\hat{p}\left(\vec{\nabla}\phi+{\phi\over h}\vec{\nabla} r\right)\right)=0,
\label{e:B.MHSga}
\end{equation}
 so taking for the integrating factor
 \begin{equation}
\vec{\nabla} \ln\phi=-{1\over h}\vec{\nabla} r,
\label{e:.MHSg}
\end{equation}
 we obtain $\vec{B}\vec{\cdot} \vec{\nabla} \hat{p} =0$, or $\hat{p}$
constant along field lines. The only solutions to Eq.
(\ref{e:.MHSg}) are the 1D $\phi=\phi(r)$ and $h= h(r)$ or $T=T(r)$,
represented in the hydrostatic relation
 \begin{equation}
{\partial \ln\phi(r)\over\partial r}=-{1\over h(r)},
\label{e:phi}
\end{equation}
 which can be integrated to give
 \begin{equation}
\phi(r)=\exp\left(-\int_{r_s}^r{{\rm d}r\over h(r)}\right),
\label{e:phiint}
\end{equation}
 where $r_s$ denotes an arbitrary base height for the radial
variations.  Thus pressure variations along arbitrarily directed
field lines follow the 1D hydrostatic scale-height function $\phi(r)$
from a constant base pressure $\hat{p}$; 3D effects are possible
because the base pressure $\hat{p}$ and the scale-height function
$\phi(r)$ or temperature profile $T(r)$ can vary from field line to
field line. One scale-height function $\phi(r)$ with temperature
profile $T(r)$ applies throughout force-free regions and in isolated
non-force-free flux tubes in thermal balance with their local
surroundings.

Substituting Eq. (\ref{e:pdef}) into the MHS Eq. (\ref{e:MHSg}) and
using Eq. (\ref{e:.MHSg}) gives the cross-field equation
 \begin{equation}
\phi(r) \vec{\nabla} \hat{p}=
{1\over4\pi}(\vec{\nabla}\vec{\times}\vec{B})\vec{\times}\vec{B},
\label{e:MHSgx}
\end{equation}
 for $\hat{p}$ allowed to vary from field line to field line.  We
adopt the single scale-height function $\phi(r)$, assuming for our
argument that one temperature profile applies throughout.  At a base
height $r_s$, $\phi(r_s) =1$, and the cross-field Eq.
(\ref{e:MHSgx}) resembles the classical MHS Eq. (\ref{e:MHS}) without
gravity.  Since the base height $r_s$ is arbitrary, it is again
evident that the classical equilibrium solutions must be applicable
in every small volume.

Since the base pressure $\hat{p}$ is constant along field lines, the
variation in the scale-height function $\phi(r)$ on the left side of
Eq. (\ref{e:MHSgx}) must be reproduced on the right side in the
Lorentz term, and so in $\vec{B}^2$ too.  Thus $\vec{B}$ must vary
like $\phi(r)^{1/2}$ along field lines in non-force-free fields
within $d/h$, which is the order of the derivatives of the
scale-height function.  Formally we can absorb $\phi(r)$ into the
magnetic field on the right side by decomposing $\vec{B}$ into the
general product
 \begin{equation}
\vec{B}=\phi_B(r)\hat{\vec{B}}=\phi_{B2}(r)^{1/2}\hat{\vec{B}}=
\phi(r)^{1/2}\zeta(r)\hat{\vec{B}}.
\label{e:Bdef}
\end{equation}
 A 1D magnetic scale-height function $\phi_B(r)$ is sufficient to
cancel $\phi(r)$; we introduce the residual radial multiplier
$\zeta(r)$ to allow differing scale-height functions, leaving
$\hat{\vec{B}}$ an unrestricted vector function. Substituting back
into Eqs. (\ref{e:MHSgx}) and (\ref{e:divB}) eliminates $\phi(r)$ and
gives
 \begin{eqnarray}
&&\hskip -.3cm\vec{\nabla}\hat{p}={1\over4\pi}\left(\left(\vec{\nabla}-{\vec{\nabla} r\over 2h(r)}\right)
\vec{\times}\zeta(r)\hat{\vec{B}}\right)\vec{\times}\zeta(r)\hat{\vec{B}},
\label{e:MHSg_hatB}\\
&&\hskip -.3cm\vec{\nabla}\vec{\cdot}\ \zeta(r)\hat{\vec{B}}=
{\vec{\nabla} r\over 2h(r)}\ \vec{\cdot}\ \zeta(r)\hat{\vec{B}}.
\label{e:div_hatB}
\end{eqnarray}
 The equations can be solved by substituting perturbative expansions
in powers of the small quantity $d/h(r)$, $\hat{p}= \sum_n{\hat{p}_n
(d/h(r))^n}$ and $\hat{\vec{B}}=\sum_n{ \hat{\vec{B}}_n (d/h(r))^n}$
for $n\geq0$ with $d$ a constant characteristic flux-tube thickness. 
The expansions separate according to powers of $(d/h(r))^n$, giving
the most-significant order-zero equations
 \begin{eqnarray}
&&\hskip -.3cm\vec{\nabla}\hat{p}_0=
{1\over4\pi}\left(\vec{\nabla}\vec{\times}\zeta(r)\hat{\vec{B}}_0\right)
\vec{\times}\zeta(r)\hat{\vec{B}}_0,
\label{e:MHSg_hatB0}\\
&&\hskip -.3cm\vec{\nabla}\vec{\cdot}\ \zeta(r)\hat{\vec{B}}_0=0.
\label{e:div_hatB0}
\end{eqnarray}

Equations (\ref{e:MHSg_hatB0}) and (\ref{e:div_hatB0}) in $\hat{p}_0$
and $\zeta(r)\hat{\vec{B}}_0$ are equivalent to the classical
plane-parallel MHS equations without gravity, which are discussed in
Appendix \ref{a:MHS}.  The solutions are the classical equilibria,
which exhibit a degenerate direction $z$ in the local coordinate
system $(x,y,z)$ taken to be arbitrarily oriented with respect to the
radial $r$.  There is no loss in generality in just taking $\zeta(r)=
1$ and $\phi_{B2}(r)= \phi(r)$ in Eq. (\ref{e:Bdef}) leaving
$\hat{\vec{B}}_0$ the unperturbed solution from Eq. (\ref{e:B2D}) and
constant in $z$.  If $z\not\perp r$ the solutions for $\vec{B}$
exhibit a variation along the field direction that goes like
$\phi(r)^{1/2}$.  In the case considered by \cite{Dungey1953a} with
$z\perp r$, distortion effects are introduced perpendicular to the
main magnetic field direction as described in the Introduction. Thus
$h_B(r)= 2h(r)$ at every $r$, where the magnetic scale height is
defined in the usual way, $h_B(r)= -[\partial \ln\phi_B(r)/\partial
r]^{-1}$.  The scale heights are coupled to most-significant order in
the perturbing gravitational field in all non-force-free MHS
equilibria containing thin flux tubes in thermal balance with their
local surroundings.

Additional perturbative equations for the base functions $\hat{p}$
and $\hat{\vec{B}}$ originate with the rescaling of the magnetic
field vector in $r$ as discussed in Appendix \ref{a:perturb}.  The
rescaling leads to the small term $\vec{\nabla}r/(2h(r))$ in both
Eqs. (\ref{e:MHSg_hatB}) and (\ref{e:div_hatB}).  The term appears on
the right side of Eq. (\ref{e:div_hatB}) and is needed since a slow
radial decrease in $|\vec{B}|$ cannot occur without a divergence of
field lines; it represents a compensating creation of flux in
$\hat{\vec{B}}(r)$ with $r$.  The term is subtracted from the curl in
the Lorentz force in Eq. (\ref{e:MHSg_hatB}) and corresponds to a
small residual pseudo-force directed between $\hat{\vec{B}}$ and the
outward radial $\vec{\nabla} r$; it represents a progressive
left-handed twist of field lines at the rate of 1 radian in every
magnetic scale height $2h(r)$.  In MHS equilibria with gravity, the
amplitude, the divergence, and the curl of $\vec{B}$ all vary on the
large scale $2h(r)$.

In principle anyway, a single thin flux tube somewhere in the
atmosphere, evacuated or in thermal balance with its surroundings,
fixes the equilibrium gas pressure for the entire atmosphere, but all
non-force-free fields must be consistent to satisfy mutually the
horizontal pressure boundary condition.  In reality of course,
magnetic fields may exhibit local deviations, and the equilibrium
state might only be reflected in the average in an atmosphere
containing many flux tubes.

A flux tube in thermal balance satisfies the condition $p_e(r)\propto
\vec{B}_e(r)^2$ whether of increased or decreased field strength
compared to the local background magnetic field.  A flux tube with
increased field strength $|\vec{B}_i(r)|>|\vec{B}_e(r)|$, must
exhibit a decreased internal gas pressure in Eq. (\ref{e:pdif}),
$p_i(r)<p_e(r)$ consistent with the extreme dark-thread evacuated
case of $p_i(r)=0$, which requires the minimum surrounding atmosphere
for a given magnetic pressure difference.  Flux tubes with less field
strength than the external magnetic field $|\vec{B}_i(r)|<
|\vec{B}_e(r)|$, must have an increased gas pressure $p_i(r)>p_e(r)$,
consistent with bright threads.

Threads appear to be isolated thin flux tubes aligned with the
background magnetic field and embedded in current sheets in a bimodal
amplitude distribution, consistent with the natural equilibrium form
discussed in Appendix \ref{a:MHSprop}. The temperature of bright
threads was indeed found to be the same as the external surroundings
within observational uncertainties (NK96). The solar EUV coronal
temperature for unresolved quiet and active regions is relatively
uniform, 1.5 -- 2.1 $\times10^6$ K (\citealt{Withbroe1975}), and a
narrow temperature range around $2\times10^6$ K is found for a broad
range of sizes and densities of X-ray loops for all but prominence
flare loops (\citealt{DavisJM+++1975, Vaiana+++1976,
Rosner+TuckerWH+Vaiana1978}). Large-area coronal temperatures derived
from the solar white-light scale height and forbidden line ratios
give temperatures somewhat lower than the EUV and X-ray temperatures,
in the range of 1 -- 2 $\times10^6$ K at the solar radial height $r=
1.15 r_\odot$ (\citealt{Guhathakurta+++1992}).

A slightly heated (or cooled) flux tube exhibits an increased (or
decreased) internal pressure $p_i(r)$ with height and thereby must
affect a decreased (or increased) internal field strength $|B_i(r)|$
or an increased (or decreased) external pressure $p_e(r)$ to satisfy
the equilibrium Eq. (\ref{e:pdif}).  Small changes in internal field
strength might be accommodated by changes in the flux-tube size as a
state of local stress in the large-scale magnetic field, and changes
in the surrounding ambient pressure $p_e(r)$ might be modified by
compensating external dynamics like we describe in Sect.
\ref{s:MHD}.  Hot flux tubes should tend to cool to the temperature
of their immediate surroundings, bringing them back to the normal MHS
equilibrium condition.  A relatively cool nonevacuated flux tube with
excess field strength tends to evacuation with height as a different
equilibrium condition, and so might just persist out of equilibrium
within about one internal scale height of its base.

Continuous 3D non-force-free equilibrium solutions are claimed
(\citealt{LowBC1985a, LowBC1991a}), but these allow an arbitrary
hydrostatic 1D gas pressure independent of the magnetic field (see
the use of $p_0(r)$ and discussion around Eq. (24) of
\citealt{LowBC1991a}, in Sect. III in \citealt{Bogdan+LowBC1986}, or
Sect. 4 in \citealt{Neukirch1997}).  In the general force balance,
that is the cross-field Eq. (\ref{e:MHSgx}) with $\phi(r)$ allowed to
vary from field line to field line, the pressure and magnetic
variations cannot be separated except in force-free regions
unconstrained by a horizontal pressure boundary condition, and so
necessarily lacking isolated thin non-force-free flux tubes evacuated
or in thermal balance with their surroundings.

We have confined our study here to the static solutions, but it is
straightforward to broaden consideration to include certain types of
steady flows.  Steady uniform flows along field lines add a flow
pressure to the parallel-field Eq. (\ref{e:B.MHSg}), which leads to a
modified hydro-steady relation in Eqs. (\ref{e:phi}) and
(\ref{e:phiint}), but the same cross-field Eq. (\ref{e:MHSgx})
results.  Thus we obtain a gas pressure containing static and wind
components coupled to the large-scale magnetic field outside flux
tubes.  For a given gas pressure variation or scale-height function
$\phi(r)$ as is defined by the magnetic field in a magnetically
dominated environment, the static contribution to the pressure scale
height $h(r)$ and corresponding temperature $kT(r)=\bar{m}g(r) h(r)$
are always larger with a steady flow than without
(\citealt{ParkerEN1960b}).  Steady flux-tube flows can modify the
coupling too.

\section{Model atmospheres}
\label{s:atm}

The derivation of an explicit static equilibrium solution is a
formidable task because such an analysis must consider the specific
form that distorted magnetic fields take in the presence of a gas. 
However an approximate coupled scale-height function can be written
 \begin{equation}
\tilde\phi(r)=\phi_a(r)+{1\over\beta_s}\phi_{B2}(r),
\label{e:phix}
\end{equation}
 where $\phi_a(r)$ and $\phi_{B2}(r)$ are nominal separate gas and
magnetic pressure solutions to the MHS Eqs. (\ref{e:MHSg}) and
(\ref{e:divB}) normalized at a base height $r_s$; the coefficient
$1/\beta_s$ defines the relative magnetic-field strength.  The
approximate coupled scale-height function $\tilde\phi(r)$ is
proportional to the total pressure $p+\vec{B}^2/(8\pi)$ of the
nominal separate solutions.  The gradient of the total pressure
approximates the gas-pressure gradient minus the Lorentz term in the
MHS equilibrium Eq. (\ref{e:MHSg}) reflecting the local balance in
Eq. (\ref{e:pdif}), as discussed at the end of Appendix
\ref{a:MHSprop}. The coupled scale-height function $\tilde\phi(r)$
thus gives a correct MHS solution where one pressure component is
dominant, $p\gg \vec{B}^2/(8\pi)$ or $\vec{B}^2/(8\pi)\gg p$.

At a base height $r_s$ taken as the stellar photospheric surface, we
assume that magnetic fields are relatively weak, $\beta_s \gg1$ and
$\tilde\phi(r) \simeq \phi_a(r)$.  There the gas pressure is dominant
and must follow its nominal atmospheric form not being much affected
by the magnetic field, whereas the magnetic field lines must be
highly distorted. The nominal atmospheric-pressure scale-height
function $\phi_a(r)$ is defined by the hydrostatic relation Eq.
(\ref{e:phiint}) with a temperature found using a transfer equation
that includes all of the usual energy input and loss mechanisms.

The atmospheric contribution $\phi_a(r)$ characteristically falls off
much more rapidly than the magnetic pressure $\phi_{B2}(r)$, so
around some transition height $r_t$ where $\phi_a(r_t)=
\phi_{B2}(r_t)/ \beta_s$, $\tilde\phi(r)$ changes from $\phi_a(r)$ to
$\phi_{B2}(r)/ \beta_s$ over about one gas-pressure scale height. 
Above $r_t$ the magnetic field is dominant $\tilde\phi(r) \simeq
\phi_{B2}(r)/ \beta_s$, and the field must follow its nominal form
irrespective of the gas pressure. There the gas pressure profile may
be very distorted from its nominal form, representing an added
atmospheric heating produced by MHS restoring flows as we discuss in
Sect. \ref{s:MHD}.

Both the magnetic and gas pressure variations must be distorted from
their nominal separate forms around the intermediate transition
height $r_t$.  Preserving the total pressure $p+\vec{B}^2/(8\pi)$
gives an approximate flux-conserving extrapolation below the
magnetically dominated corona and an ostensible gas-pressure
extrapolation above the photosphere and chromosphere.  The total
relative pressure around the intermediate transition height goes from
$A_{\rm tot}(\phi_a(r)+ \phi_{B2}(r)/ \beta_s)$ for a nominal
superposition of independent gas and magnetic pressures $\phi_a(r)$
and $\phi_{B2}(r)/ \beta_s$ with $A_{\rm tot}$ the total surface
area, to $(A_{\rm tot}-A_{\rm vac}) \tilde\phi(r)+ A_B\tilde\phi(r)$
for the coupled pressure $\tilde\phi(r)$ with $A_{\rm vac}$ the
evacuated area and $A_B$ the magnetically filled area.  The two
pressure totals are equal in general only with $\tilde\phi(r)$ from
Eq. (\ref{e:phix}) and when the field-filling regions are evacuated
with $A_B=A_{\rm vac}$.  Anyway the choice of coupled scale-height
function is not too critical for our demonstration, as smooth
switching functions with correct asymptotic behavior in the
photosphere and corona exhibit similar temperature profiles even
around the transition height.

For demonstration purposes, we take $\phi_a(r)$ for a polytrope
atmosphere as derived in Appendix \ref{a:poly}
 \begin{equation}
\phi_a(r)=\left(1-{\Gamma-1\over\Gamma}{r_s\over h_s}
\left(1-{r_s\over r}\right)\right)^{\Gamma\over\Gamma-1},
\label{e:phia}
\end{equation}
 where $h_s\equiv h(r_s)= kT_s/ (\bar{m}g_s)$ denotes the surface
scale height and $\Gamma$ the ratio of specific heats or polytrope
adiabat.

A nominal scale-height function $\phi_{B2}(r)$ is defined for a
single-spherical-harmonic potential magnetic field. The potential
magnetic form seems to be consistent with what is observed in the
lower solar corona for $r<1.6 r_\odot$
(\citealt{Altschuler+Newkirk1969, Schatten+Wilcox+Ness1969}). The
potential magnetic field vector is the gradient of a sum of scalar
spherical-harmonic component functions, each of which is the
separable product of a 2D surface function and radial multiplier $1/
r^{\ell+1}$.  The resulting magnetic field in each spherical harmonic
goes like $1/r^{\ell+2}$ in all its vector elements; a monopole
$\ell=0$ exhibits a $1/r^2$ radial falloff, a dipole field $\ell=1$,
a $1/r^3$ falloff, etc.  For a single spherical harmonic $\ell$, the
magnetic energy density falls off like
 \begin{equation}
\phi_{B2}(r)=\left({r_s\over r}\right)^{2\ell+4}.
\label{e:phiB2}
\end{equation}
 With a superposition of spherical harmonics, a nonuniform radial
dependence can occur, and a representative global-average spherical
harmonic $\ell$ might be obtained by appropriately weighting
flux-tube locations.

The temperature is written from the hydrostatic Eq. (\ref{e:phi})
 \begin{equation}
kT(r)=\bar{m}g(r)h(r)=
-\bar{m}g(r)\left({1\over\phi(r)}{\partial\phi(r)\over\partial r}\right)^{-1}.
\label{e:T}
\end{equation}
 Taking $\phi(r)$ to be the approximate $\tilde\phi(r)$ from Eq.
(\ref{e:phix}) with $\phi_a(r)$ from Eq. (\ref{e:phia}) and
$\phi_{B2}(r)$ from Eq. (\ref{e:phiB2}), we obtain an explicit
formula for $T(r)$. In the corona $\phi(r)\simeq \phi_{B2}(r)/
\beta_s$, and the formula exhibits the limiting scale height
 \begin{equation}
h_{\rm cor}(r)={r\over 2\ell+4},
\label{e:hcor}
\end{equation}
 and temperature
 \begin{equation}
k T_{\rm cor}(r)= {GM_s\bar{m}\over (2\ell+4) r} = {\bar{m}g(r)r\over 2\ell+4}.
\label{e:Tcor}
\end{equation}
 The limiting base coronal temperature $T_{\rm cor}(r_s)$ is defined
entirely by the stellar surface gravity $g_s$, radius $r_s$, mean
particle mass $\bar{m}$, and main magnetic spherical harmonic
$\ell$.  For a toroidal field $\ell=2$, the scale height at the
stellar surface is $h_{\rm cor}(r_s)= r_s/8$, which gives a solar
base coronal temperature of $T_{\rm cor}(r_s)= 1.73\times 10^6$ K
(using $\bar{m}= 0.6 m_p$ for $m_p$ the proton mass, $\bar{m}=1\times
10^{-24}$ g, $r_s=r_\odot= 6.96\times 10^{10}$ cm, $g_s= 2.75\times
10^4$ cm s$^{-2}$, and $k=1.38\times 10^{-16}$ ergs K$^{-1}$).  With
a solar base coronal temperature of $T_{\rm cor}(r_s)=1.6\times 10^6$
K, we obtain the spherical harmonic $\ell=2.33$.

\begin{figure}
\centering
\includegraphics[width=8.7cm]{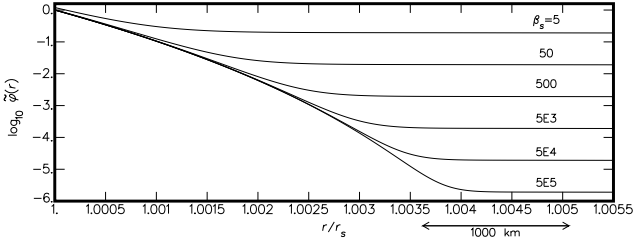}
 \caption{Log of scale-height function $\log_{10}\tilde\phi(r)$ for a
solar polytrope atmosphere with $\Gamma=1.1$ and a potential magnetic
field $\ell=2.33$ for different $\beta_s$.}
 \label{f:phibet}
\end{figure}

Figure \ref{f:phibet} shows some example scale-height functions
$\tilde\phi(r)$ from Eqs. (\ref{e:phix}) -- (\ref{e:phiB2}) in the
lowest part of the solar atmosphere around the transition height
$r_t$. Scale-height functions with the polytrope adiabat $\Gamma=1.1$
and spherical harmonic $\ell=2.33$ are plotted for different 
$\beta_s$, using the solar parameters and surface photospheric
temperature $T_s=$6820 K from the VAL model atmospheres
(\citealt{VALIII}), which defines the surface scale height
$h_s=kT_s/(\bar{m}g_s)$ for $\phi_a(r)$ in Eq. (\ref{e:phia}). As all
of the curves are based on the same $\Gamma$, they coincide until the
transition height $r_t$ for the model is reached, and then switch
rapidly to the nominal magnetic $\phi_{B2}(r)$, which appears to be
relatively flat on the log scale.

\begin{figure}
\centering
\includegraphics[width=8.7cm]{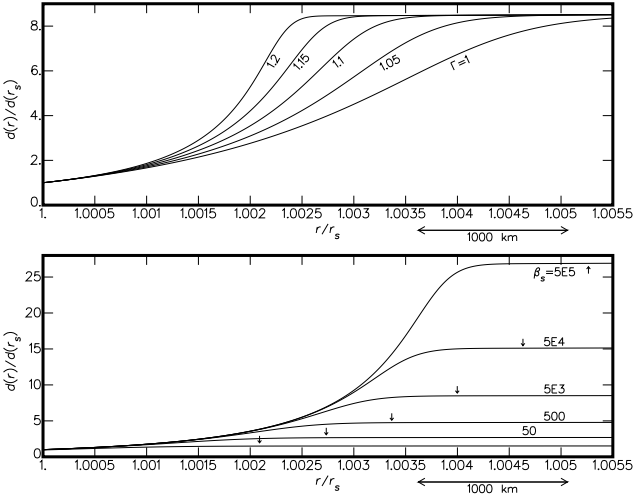}
 \caption{Relative flux-tube diameter $d(r)/d(r_s)$ in lower solar
atmosphere with $\ell=$2.33 and $\beta_s=$5000 for different
polytrope adiabats $\Gamma$ ({\it upper panel}), and with $\ell=$2.33
and $\Gamma=1.1$ for different $\beta_s=$ 5, 50, $\dots$ 5.E5 ({\it
lower panel}).  The arrow on each curve locates the most rapid
temperature change for the model.}
 \label{f:fluxtubes}
\end{figure}

The radial expansion of a flux tube is another way to visualize the
field strength decrease in the atmosphere and the properties of the
coupled scale-height function.  For conservation of the total flux
through the cross-sectional area of a flux tube $\pi (d/2)^2
|\vec{B}|$, the flux-tube diameter $d(r)$ must increase with radial
distance $r$, $d(r) \propto |\vec{B}|^{-1/2} \propto
\phi(r)^{-1/4}$.  Figure \ref{f:fluxtubes} portrays the relative
flux-tube diameter as a function of height $d(r)/d(r_s)$ for
different solar atmospheric models denoted by $\Gamma$ and
$\beta_s$.  

For a given polytrope adiabat $\Gamma$, all flux tubes exhibit
approximately the same relative shape up to a height that depends
upon $\beta_s$, where the field lines straighten up.  Stronger fields
exist with a lower transition height and exhibit less overall
relative expansion before straightening up.  The canopy depends upon
the surface distribution of the fields, but strong fields might give
the appearance of a lower canopy height too consistent with 
observations of sunspots (\citealt{Giovanelli+JonesHP1982}). The
straightening height is somewhat below the location of the fastest
temperature change due to the differing dependencies: the field
strength, which goes like $\phi(r)^{1/2}$, corrects from its
atmospherically determined form at $r_t$, but the most rapid
temperature change occurs significantly higher where the slope of the
scale-height function flattens out, as $T(r)\propto -(\partial
\ln\phi(r) /\partial r)^{-1}$.

\begin{figure}
\includegraphics[width=8.7cm]{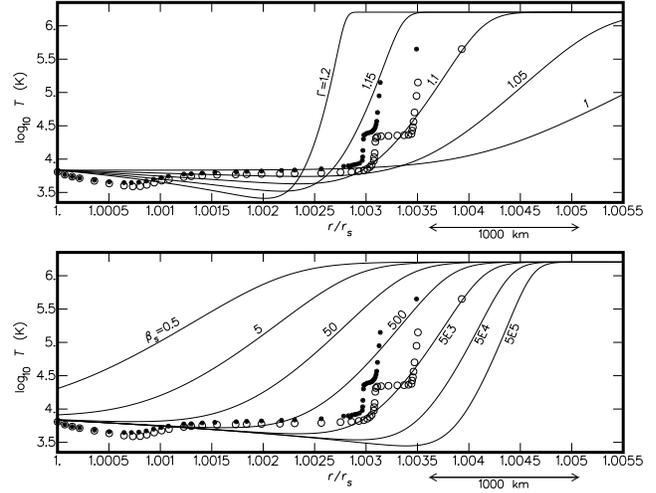}
 \caption{Lower solar atmosphere $\log_{10}T(r)$ with $\ell=2.33$ and
$\beta_s=$5000 for different polytrope adiabats $\Gamma$ ({\it upper
panel}), and with $\ell=2.33$ and $\Gamma=1.1$ for different
$\beta_s$ ({\it lower panel}), shown with the solar VAL A ({\it open
circles}) and F ({\it filled circles}) atmospheres.}
 \label{f:transreg}
\end{figure}

Figure \ref{f:transreg} illustrates the temperature variation in the
lower solar atmosphere using Eq. (\ref{e:T}) with $T_s=$6820 K from
the VAL atmospheres.  The upper panel shows the model atmospheres
with $\ell=2.33$ and $\beta_s=$5000 for different polytrope adiabats
$\Gamma$, and the lower panel the model atmospheres with $\ell=2.33$
and $\Gamma= 1.1$ for different $\beta_s$.  The VAL A and F model
temperatures for relatively cool inner network and hot network bright
points are also shown.

In the solar photosphere, the nominal atmospheric polytrope follows
the VAL temperature roughly, giving the best tradeoff between 
photospheric and chromospheric temperatures with $\Gamma\simeq 1.1$. 
Of course the features of a real solar atmosphere can never be well
approximated by a polytrope. The shape of the temperature function
through the chromosphere and transition height, where the
scale-height function $\tilde\phi(r)$ goes from $\phi_a(r)$ to
$\phi_{B2}(r)/\beta_s$, depends upon $\Gamma$ and $\beta_s$ but not
on $\ell$, which separately determines the base coronal temperature. 
Temperature profiles with $\beta_s\simeq 5000$ or 1000 give best
agreement around the transition height with the VAL A or F
atmospheres, respectively, averaging approximately through the
Ly$\alpha$ plateau.

For flux-tube evacuation in the photosphere $\vec{B}(r_s)^2 = 8\pi
p(r_s)$, and we obtain a maximum field strength of $|\vec{B}(r_s)|=$
1715 G using $p(r_s)=1.17\times 10^5$ dynes cm$^{-2}$ from the VAL
solar models; $1/\beta_s^{1/2}$ might be taken as the fractional area
covered by evacuated magnetic fields. Strictly the coefficient
$1/\beta_s$ represents the coronal base gas or non-force-free
magnetic pressure extrapolated back to $r_s$, which can be no larger
than the average surface magnetic pressure.  Thus the fractional area
covered by evacuated magnetic fields at $r_s$ must be at least $1/
\beta_s^{1/2}$, or $1/71$ for $\beta_s= 5000$ and $1/32$ for
$\beta_s= 1000$, which corresponds to a minimum base coronal field
strength of $|\vec{B}_{\rm cor}|=$ 24 G for $\beta_s= 5000$ and
$|\vec{B}_{\rm cor}|=$ 54 G for $\beta_s= 1000$, within the range of
solar observations (\citealt{LinH+Penn+Tomczyk2000}).  A lower
transition region occurs where the photospheric flux-tube areal
coverage and coronal magnetic field strength are larger near active
regions, consistent with the known tendency.

The solar atmospheric model of \cite{Fontenla+Avrett+Loeser1990},
which includes particle diffusion and conduction effects, lacks the
Ly$\alpha$ plateau and exhibits a much more abrupt transition region
than what we obtain.  While radiative, diffusive, and conductive
losses must be largely balanced by MHS restoring flows in the corona,
as we discuss in Sect. \ref{s:MHD}, loss mechanisms can influence the
detailed shape of transition-region profiles. Real radiative models
coupled to more general magneto-hydro-steady equilibrium solutions
need to be developed, but such work may be complicated by the
intrinsic limitations and uncertainties in atmospheric modeling,
especially associated with inhomogeneous magnetic structure
(\citealt{Ayres1981, Carlsson+SteinRF1995}).  It is possible that
general methods based upon transition-region emissivity profiles,
which have been used to determine the nonradiative atmospheric
heating contribution (\citealt{CraigI+BrownJC1976a,
AndersonSW+RaymondJC+vanBallegooijen1996}), might be able to
distinguish coronal heating by MHS restoring flows from other heating
mechanisms.

\begin{figure}
\includegraphics[width=8.7cm]{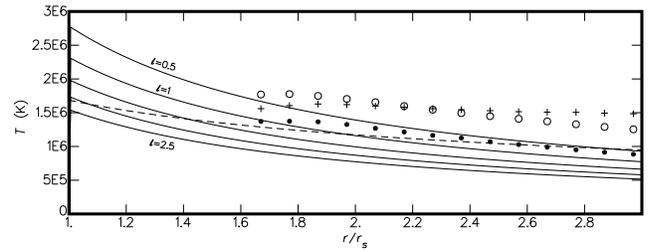}
 \caption{Static solar corona $T(r)$ for $\ell=$.5, 1, 1.5, 2, and
2.5 (from top to bottom) shown with temperature from white-light
intensity in the quiet equator ({\it filled circles}), in SW ({\it
crosses}), and in SE ({\it open circles}) streamers
(\citealt{Newkirk+Dupree+Schmahl1970}), and from line-width
measurements ({\it dashed line}) (\citealt{Jarrett+vonKluber1958}).}
 \label{f:farcor}
\end{figure}

Figure \ref{f:farcor} illustrates the solar atmospheric temperature
function $T(r)$ for various values of $\ell$ with the coronal
temperatures inferred from the white-light-intensity radial gradient
taken from eclipse photographs (\citealt{Newkirk+Dupree+Schmahl1970})
and inferred from FeXIV 5303\AA\ line-width measurements
(\citealt{Jarrett+vonKluber1958}).  The limited resolution in $r$ in
the figure hides the transition region near $r_s$.  The coronal model
for the spherical harmonic $\ell=1$ seems to give the best overall
agreement with the measured radial profiles, but $\ell\simeq2.3$
better matches the base coronal temperature. However all of the model
curves show a more rapid falloff than the measured temperature
profiles. Models containing an outward wind should give
systematically higher temperatures with height in the corona.

\section{MHS restoring flows}
\label{s:MHD}

Under usual conditions, e.g. quasi-steady, the large-scale and
long-term system average is represented by the equilibrium state.  In
the quasi-steady approximation, the system evolution can be described
by a sequence of nearby equilibria.  \cite{LowBC1977b} shows that the
quasi-steady approximation is applicable when magnetic adjustments
are fast compared to evolutionary processes and the magnetic field is
dominant over the gas, as in the lower solar corona.

A qualitative feature of MHD processes in a slowly evolving dominant
magnetic field where reconnections are fast is an atmospheric
restoring flow.  Hydro-steady equilibria are always self-restoring. 
In the absence of sufficient external gas or flow pressure, a
time-dependent outward gas acceleration arises outside and along
rigidly held non-force-free flux tubes in the overall MHD balance, as
described in Appendix \ref{a:MHD}. Flows in a fixed magnetic field act 
generally to reestablish the equilibrium atmosphere, heating the gas 
compressively.  As long as the timescale for magnetic evolution is 
relatively long, such MHD processes can have a large-scale effect
since the pressure perturbation that drives the flow propagates
horizontally away from flux tubes at the sound speed.  

The available energy for outflows must be limited and the atmosphere
has to eventually cool, distorting and geometrically stressing the
coupled magnetic field, which contributes to magnetic instability. 
The magnetic field can be perturbed in other ways too, for example by
new flux coming up from the convection zone below or by photospheric
twisting and flux-tube motions.  Magnetic evolution is complicated
since unstable equilibria are generally possible where perturbations
{\it reduce} the overall energy of the system. Also local deviations
may persist within field allowances before equilibria can be
resolved. However if the new magnetic equilibrium after fast
reconnection is mostly force-free containing rigidly held
non-force-free flux tubes, outflows again must reestablish the
equilibrium gas-pressure profile of a depleted cool atmosphere,
applying the argument from Appendix \ref{a:MHD}.

MHS restoring flows may provide an important coronal heating
mechanism, which can exist consistently along with other heating
processes.  By offseting the total coronal energy requirement, other
heating sources should add stability to the large-scale magnetic
field and thus longevity to the magnetic evolution.  The total
coronal energy requirement is estimated conventionally using
heat-flux scaling relations for loops and homogeneous regions
(\citealt{Rosner+TuckerWH+Vaiana1978, Hearn+Kuin1981,
HammerR1982b}).  If sufficient magnetic energy is not available from
evolutionary processes to offset losses, then pure force-free cool
equilibria may be the only possibility.

Localized heating in flux tubes, which is thought to be an important
coronal heating process (\citealt{Vaiana+Rosner1978,
Airapetian+Smartt1995}), may affect ambient MHS restoring flows. 
Heated flux tubes must exhibit field-line perturbations that stress
the large-scale magnetic field and tend to produce outflows in the
surroundings at least on the average, which heat the ambient
atmosphere above the normal equilibrium temperature.  The ambient
hydrostatic imbalance can have a relatively large scale of influence
since it propagates horizontally away from its source much faster
than thermal diffusion effects. Such post-reconnection hydrodynamic
heating acting throughout the coronal volume may help overcome the
often-cited difficulty that reconnection events are so spatially
localized and short-lived that only a larger event rate than what is
directly inferred from observations could account for coronal heating
(\citealt{Heyvaerts+Priest1984, Hudson1991a}).

Thus MHD processes may lead to refilling outflows, a `magnetic
suction', which might be in evidence as the large-scale flows seen in
the vicinity of coronal voids
(\citealt{Wagner+Newkirk+SchmidtHU1983}), or be responsible for the
supergranular-scale chromospheric upward velocities interpreted from
post-flare solar spectral-line blue shifts
(\citealt{Schmieder+++1987, Cauzzi+++1996}), or appear as spicule
eruptions in the chromosphere. The upward mass flux produced by solar
spicule events is about 100 times the total solar-wind mass flux, but
represents only a small fraction of its total energy flux
(\citealt{Pneuman+KoppRA1978}).

The remarkable feature of an MHS corona is that the temperature is
essentially a geometric parameter defined by the stellar surface
gravity $g_s$, radius $r_s$, mean particle mass $\bar{m}$, and
magnetic spherical harmonic $\ell$
 \begin{equation}
k T_{\rm cor}(r_s)={\bar{m}g_s r_s\over 2\ell+4}.
\label{e:Tcorodot}
\end{equation}
 The observed average solar base coronal temperature of $1.6\times
10^6$ K corresponds to a spherical harmonic of $\ell=2.33$ consistent
with the large-scale solar magnetic field, which exhibits substantial
power in the toroidal $\ell=2$ and spherical harmonics $\ell=3$ or 4
as evidenced by the presence of active longitudes. With a general
mixture of spherical harmonics, discrepant horizontal gas-pressure
boundary conditions at flux tubes can arise, and a nonuniform
large-scale surface temperature distribution is possible as a
deviation from the pure MHS equilibrium state.

Formulae for the average base coronal temperature like Eq.
(\ref{e:Tcorodot}) have been suggested already in the hypothesis of a
`geometric boundary condition', relevant for different kinds of
heating models where the base pressure scale height is proportional
to the stellar radius (\citealt{Menzel1968a, Scudder1992a}).  The
relation appears to give reasonable coronal temperatures for a range
of stellar types (\citealt{Williams+Mullan1996}).

\begin{acknowledgements}
 The author is grateful to many for useful discussion during the
course of this work, especially to Eric Priest and Ray Smartt for
their careful reviews of the manuscript, and to the referee for many
insightful remarks and helpful suggestions.
 \end{acknowledgements}

\appendix
\section{Local MHS equilibria}
\label{a:MHS}

There are well-known solutions to the classical static problem
describing a gas existing in conjunction with elongated magnetic
fields without a perturbing gravitational field (\citealt[Chapter
6]{ParkerEN1979}).  It is helpful to revisit the classical problem
here in a general way.  The classical MHS equilibrium equations
without gravity are the static equilibrium equation
 \begin{equation}
\vec{\nabla} p={1\over4\pi}(\vec{\nabla}\vec{\times}\vec{B})\vec{\times}\vec{B},
\label{e:MHS}
\end{equation}
 and Gauss' Law, Eq. (\ref{e:divB}).

Plane-parallel solutions can be developed based upon the general
Cartesian vector function
 \begin{equation}
\vec{B}= \left({\partial a(\vec{x})\over\partial y},
-{\partial a(\vec{x})\over\partial x}-{\partial b(\vec{x})\over\partial z},
{\partial b(\vec{x})\over\partial y}\right).
\label{e:CartB}
\end{equation}
 The $x$ and $z$ vector elements of $\vec{B}$, $\partial a/\partial
y$ and $\partial b/\partial y$, are taken to be arbitrary functions,
and the $y$ element is written so that Gauss' Law is always
satisfied.  It is convenient to use derivative forms in the vector
elements to avoid complicating integrals in the expression.

Any solutions to Eq. (\ref{e:MHS}) must satisfy the two
conditions $\vec{B}\vec{\cdot} \vec{\nabla} p=0$ and
$(\vec{\nabla}\vec{\times}\vec{B})\vec{\cdot} \vec{\nabla} p=0$,
which are written with $\vec{B}$ from Eq. (\ref{e:CartB})
 \begin{eqnarray}
&&\hskip -.3cm{\rm C}_{xy}a-{\rm C}_{yz}b=0,
\label{e:B.Dp}\\
&&\hskip -.3cm\left({\rm C}_{xz}{\partial\over\partial x}+{\rm C}_{yz}{\partial\over\partial y}\right)a+
\left({\rm C}_{xy}{\partial\over\partial y}+{\rm C}_{xz}{\partial\over\partial z}\right) b=0,
\label{e:curlB.Dp}
\end{eqnarray}
 using the linear commutation operators ${\rm C}_{xy}\equiv{\partial
p\over\partial x}{\partial\over\partial y} -{\partial p\over\partial
y} {\partial\over\partial x}$, etc., which contain $p(\vec{x})$ as an
implicit function.

The two Eqs. (\ref{e:B.Dp}) and (\ref{e:curlB.Dp}) can be used to
write two of the functions in terms of the third, for example
$a(\vec{x})$ and $b(\vec{x})$ in terms of $p(\vec{x})$.  The third
relation, Eq. (\ref{e:MHS}) in the vector direction perpendicular to
both $\vec{B}$ and $\vec{\nabla}\vec{\times}\vec{B}$, gives a
differential relation for the third quantity.  The separate vanishing
of every commutator term is a reduction that is satisfied with
certain general functional dependencies between the variables and
leads to a well-known self-consistent form for the third relation,
the Grad-Shafranov Equation.  Other possibilities are not considered
here.

Taking every commutator term to vanish in Eqs. (\ref{e:B.Dp}) and
(\ref{e:curlB.Dp}), but without other reduction, gives $a(\vec{x})$,
$b(\vec{x})$, and spatial derivatives of $a(\vec{x})$ and
$b(\vec{x})$ as functions of $p(\vec{x})$ alone, which is only
possible with specific spatial forms for real functions, so other
reductions are required in general.  Taking $\partial/\partial z= 0$
gives a consistent solution defined by the functional dependencies
$a=a(p)$ in Eq. (\ref{e:B.Dp}) and ${\partial b\over\partial y}=
{\partial b\over\partial y}(p)$ in Eq. (\ref{e:curlB.Dp}).  A similar
solution is obtained by taking $\partial/\partial x= 0$.  These two
solutions or linear combinations are equivalent with rotation of the
$x$ -- $z$ axes around $y$.  Taking $z$ as the degenerate direction
gives the usual 2D solution, ordinarily written with $a$ as the
implicit function
 \begin{equation}
\vec{B}= \left({\partial a(x,y)\over\partial y},
-{\partial a(x,y)\over\partial x},B_z(a(x,y))\right).
\label{e:B2D}
\end{equation}
 The degenerate $z$ is a direction of field elongation.  The magnetic
potential function $a(x,y)$ defines a 2D planform for the solutions,
which determines the spatial variations in the longitudinal field
element $B_z=B_z(a(x,y))$ and gas pressure $p=p(a(x,y))$ as arbitrary
1D mappings.  The magnetic field is everywhere perpendicular to the
gradient of $a(x,y)$, since $\vec{B}\vec{\cdot} \vec{\nabla} a(x,y)
=0$.

It is widely believed that this 2D solution is the only 
plane-parallel one.  Using perturbative expansions, Parker 
(\citeyear[Sect. 14.2]{ParkerEN1979}) shows that no nearby 3D
solutions exist for bounded quantities in an infinite spatial
domain.  However 3D variations do arise as large-scale deviations
from plane parallel (\citealt{Arendt+Schindler1988}). In the common
shorthand,`2.5D' solutions refer to perturbed solutions in the
problem with gravity, which admit large-scale variations in the
degenerate direction of the magnetic field.

\section{Properties of local MHS equilibria}
\label{a:MHSprop}

Substituting $\vec{B}$ from Eq. (\ref{e:B2D}) into Eq. (\ref{e:MHS})
gives the governing equation for the implicit magnetic potential
function $a(x,y)$ for the static equilibrium problem without gravity
 \begin{equation}
\left(\vec{\nabla}^2 a+{d\over da}\left(4\pi p(a)+
{B_z(a)^2\over2}\right)\right)\vec{\nabla} a=0.
\label{e:GS+}
\end{equation}
 Where $\vec{\nabla} a\not=0$, the potential function must satisfy the
Grad-Shafranov (GS) Equation, written
 \begin{equation}
\vec{\nabla}^2 a(x,y)=-P'(a(x,y)),
\label{e:GS}
\end{equation}
 defining the total pressure $P(a)= 4\pi p(a)+ B_z(a)^2/2$, where
$P'(a)=dP(a)/da$ .

The total pressure $P(a)$ and its component functions $p(a)$ and
$B_z(a)$ are all one dimensional and nonlinear, but must all be
well-defined everywhere in the solution domain consistent with their
forms at the boundaries. Strictly, disagreeing $z$ boundary
conditions are inconsistent with the $z$ independence of the
solutions; differences might produce small deviations in the
solutions like twist or divergence or be a source of dynamical
instability.  It is popular to restrict consideration to entirely
force-free solutions with $p'(a)=0$; the restriction does not change
the nature of the basic GS Eq. (\ref{e:GS}) for $a(x,y)$ but requires
the specific current density $\vec{J}= {c\over4\pi} B_z'(a) \vec{B}$
for an arbitrary $B_z(a)$, as is evident by expanding $\vec{J}\equiv
{c\over4\pi} \vec{\nabla}\vec{\times}\vec{B}$ using $\vec{B}$ from
Eq. (\ref{e:B2D}) with $\vec{J}\propto \vec{B}$.

The magnetic potential function $a$ can be seen to be constant
everywhere in the local plane of $\vec{B}$ and $\vec{J}$.  From Eq.
(\ref{e:MHS}), $\vec{J}\vec{\cdot} \vec{\nabla} p(a) = 0$.  Then
$\vec{J}\vec{\cdot} \vec{\nabla} p(a) = (\vec{J}\vec{\cdot}
\vec{\nabla} a)\ p'(a) =0$. Thus $\vec{J}\vec{\cdot} \vec{\nabla} a
=0$ at least when $p'(a)\not=0$, and when $p'(a)=0$, $\vec{J}
\parallel \vec{B}$ so $\vec{J}\vec{\cdot} \vec{\nabla} a =0$ anyway.
We think of $a$ as constant on ribbon-like current sheets, the
solution surfaces being more curved along $\vec{J}$ as it is defined
by the derivatives of $\vec{B}$.

The GS Eq. (\ref{e:GS}) has the unusual feature that it contains the
1D filter function $P'(a)$.  The Fourier transform of the GS Equation
in the linear case, with $P'(a)= k_0^2 a$ for $k_0$ a constant
wavenumber, requires a transform function $\bar{a}(k_x,k_y)$ that is
zero everywhere in its 2D wavenumber domain $\vec{k}= (k_x, k_y)$
except on a thin annulus at $|\vec{k}|= k_0$, where arbitrary complex
values Hermitian in $\pm\vec{k}$ are allowed.  Taking the Fourier
transform back gives the potential function $a(x,y)$ as a common
Bessel-function radial kernel times azimuthal factor convolved with a
spatial distribution of delta-function source points. Boundary
conditions on $a(x,y)$ constrain the azimuthal factor, leading to
possible planar solutions defined by all power at one azimuth, or
axisymmetric solutions with power uniformly distributed in azimuth,
consistent with our visualization of current sheets and cylindrical
flux tubes.  Physical arguments show that nonlinear kernels are
similarly constrained by boundary conditions
(\citealt{Vainshtein+ParkerEN1986}).

Random spatial distributions of a common GS kernel are nonlinear GS
solutions too.  We take the potential function $a(x,y)$ to be the
convolution of a distribution of sources $D(x,y)$ with a common 2D
kernel function $A(x,y)$, $a(x,y)= D(x,y)* A(x,y)$. The distribution
is written $D(x,y)=\sum_j c_j \delta(x-x_j,y-y_j)$ counting sources
$j$ of varying strength $c_j$, where $\delta(x,y)$ denotes the 2D
delta function.  For a spatially incoherent or random distribution
$D$ of equal strength sources with all $c_j=1$, $D$ raised to a power
is $D$ alone; for equal amplitude sources with $c_j=\pm 1$, $D$
raised to an odd power is $D$ alone. For an analytic nonlinear driver
function $P(a)$ the convolution factors out $P'(a)= P'(D*A)=
D*P'(A)$, when the distribution in $D$ is incoherent, disjoint, and
contains suitably restricted amplitudes.  Then the GS Eq.
(\ref{e:GS}) reduces to the same GS Equation, but for the common
kernel function $A$ in place of $a$. Dark and bright threads appear
to be isolated thin flux tubes aligned with the background magnetic
field organized in current sheets suggestive of a bimodal amplitude
distribution of a common flux-tube kernel.

\cite{ParkerEN1972} considers MHS equilibria that contain a 
relatively strong constant background field.  Expanding the Lorentz
term in Eq. (\ref{e:MHS}) gives an alternate form for the MHS
equation
 \begin{equation}
\vec{\nabla}\left(p+{\vec{B}^2\over8\pi}\right)=
{1\over4\pi}\left(\vec{B}\vec{\cdot}\vec{\nabla}\right)\vec{B},
\label{e:MHSalt}
\end{equation}
 or with the superposition $\vec{B}(\vec{x}) =\vec{B}_e+
\vec{B}_\sim(\vec{x})$ for a constant background field $\vec{B}_e$
and a spatially varying field $\vec{B}_\sim(\vec{x})$, we obtain
 \begin{equation}
\vec{\nabla}\left(p+{\vec{B}^2\over8\pi}\right)=
{1\over4\pi}\left(\vec{B}_e\vec{\cdot}\vec{\nabla}\right)\vec{B}_\sim+
{1\over4\pi}\left(\vec{B}_\sim\vec{\cdot}\vec{\nabla}\right)\vec{B}_\sim.
\label{e:MHSalta}
\end{equation}
 Taking the divergence, applying Gauss's Law $\vec{\nabla}
\vec{\cdot} \vec{B}_\sim=0$, and using the property that the total
pressure $p+\vec{B}^2/(8\pi)$ is a bounded quantity in an infinite
domain, we obtain
 \begin{equation}
\left(\vec{B}_e\vec{\cdot}\vec{\nabla}\right)\vec{B}_\sim=0,
\label{e:P72a}
\end{equation}
 at least to first order in the small quantity $|\vec{B}_\sim|/
|\vec{B}_e|$.  Thus we have
 \begin{equation}
\vec{\nabla}\left(p+{\vec{B}^2\over8\pi}\right)\simeq 0.
\label{e:P72b}
\end{equation}
 with equality to first order in $|\vec{B}_\sim|/ |\vec{B}_e|$. Eq.
(\ref{e:P72a}) proves to be accurate to all orders in the
perturbative expansions, and the equilibrium Eqs. (\ref{e:pdif}) and
(\ref{e:P72}) are justified at least in the presence of a relatively
strong constant background field, where the internal flux-tube field
is defined $\vec{B}_i\equiv \vec{B}_e+ \vec{B}_\sim (=\vec{B})$.
Additional discussion on Parker's perturbative expansion is contained
in Appendix \ref{a:perturb}.

The Parker theorem Eq. (\ref{e:P72a}) reflects the special features
of the classical equilibria in Eq. (\ref{e:B2D}) in the degenerate
direction of the magnetic field $z$.  MHS equilibria allow only
certain alignments for the background field.  A constant magnetic
field can enter into the $z$ vector element $B_z(a)$ without
producing other ramifications, whereas an added offset in the $x$ or
$y$ elements of $\vec{B}$ requires adding a uniformly inclined plane
to $a(x,y)$.  Such a plane limits the solutions and thus cannot be
considered general, and even appears to be precluded by the
boundedness of $a(x,y)$: low-wavenumber components in the Fourier
domain needed to represent the added plane are at odds with the
Fourier ring solutions in the linear case of the GS Eq.
(\ref{e:GS}), and also contrary to the form of a superposition of
spatially compact axisymmetric nonlinear kernels.

\section{Perturbative expansions}
\label{a:perturb}

The less-significant order $n>0$ equations in the perturbative series
developed from Eqs. (\ref{e:MHSg_hatB}) and (\ref{e:div_hatB}) are
written for $\zeta(r)=1$; from the MHS Eq. (\ref{e:MHSg_hatB})
 \begin{equation}
\begin{array}{l}
\vec{\nabla} \hat{p}_1
-{1\over4\pi}\left((\vec{\nabla}\vec{\times}\hat{\vec{B}}_0)\vec{\times}\hat{\vec{B}}_1
+(\vec{\nabla}\vec{\times}\hat{\vec{B}}_1)\vec{\times}\hat{\vec{B}}_0\right)=\\
\quad{1\over8\pi d}(\vec{\nabla} r\vec{\times}\hat{\vec{B}}_0)\vec{\times}\hat{\vec{B}}_0,
\label{e:MHSg_hatB1}
\end{array}
\end{equation}
 or for the general order $n>0$
 \begin{equation}
\begin{array}{l}
\vec{\nabla} \hat{p}_n -{1\over4\pi}\sum_{j=0}^n{
(\vec{\nabla}\vec{\times}\hat{\vec{B}}_j)\vec{\times}\hat{\vec{B}}_{n-j}}=\\
\quad{1\over8\pi d}\sum_{j=0}^{n-1}{
(\vec{\nabla} r\vec{\times}\hat{\vec{B}}_j)\vec{\times}\hat{\vec{B}}_{n-j-1}
\left(1+2j{\partial h(r)\over\partial r}\right)}.
\label{e:MHSg_hatBn}
\end{array}
\end{equation}
 From Eq. (\ref{e:div_hatB})
 \begin{equation}
\vec{\nabla}\vec{\cdot}\hat{\vec{B}}_1=-{1\over2d}(\vec{\nabla} r\vec{\cdot}\hat{\vec{B}}_0),
\label{e:div_hatB1}
\end{equation}
 or in general for $n>0$
 \begin{equation}
\vec{\nabla}\vec{\cdot}\hat{\vec{B}}_n=-{1\over2d}(\vec{\nabla} r\vec{\cdot}\hat{\vec{B}}_{n-1})
\left(1+2(n-1){\partial h(r)\over\partial r}\right).
\label{e:div_hatBn}
\end{equation}
 The equations are linear in the variables $\hat{p}_n$ and
$\hat{\vec{B}}_n$ based upon order $<n$ variables. The order-zero
base functions $\hat{p}_0$ and $\hat{\vec{B}}_0$ are 2D and constant
in the degenerate magnetic field direction $z$; 3D effects can be
introduced with the cross-product terms or with general variation in
the function $h(r)$ with $r$ not perpendicular to $z$. For
consistency in Eqs. (\ref{e:MHSg_hatB}) and (\ref{e:div_hatB}),
$\hat{\vec{B}}_0 \vec{\cdot} \vec{\nabla} \hat{p}_0 =0$, which is the
order-zero component of $\vec{B}\vec{\cdot} \vec{\nabla} \hat{p} =0$.

A wide latitude is allowed for choosing perturbative quantities, and
as long as they are relatively small almost everywhere the standard
procedure remains valid.  Van Ballegooijen (\citeyear[Appendix
A]{vanBallegooijen1985}) questions the \cite{ParkerEN1972} choice of
perturbative quantity and equation separation procedure, but van
Ballegooijen's assigning of approximate magnitudes to individual
terms to separate his equations is not an algebraic procedure, and
his resulting equations have inherent contradictions as described by
\cite{ParkerEN1987a}. \cite{ParkerEN1972} does not ascribe a
particular functional meaning to his kernel $\epsilon$, quoting from
the beginning of his Sect.  II: ``Expand the field $\vec{b}$ and
pressure $p$ in ascending powers of {\it some} parameter $\epsilon$,
which is of the order of $\epsilon= |\vec{b}|/|\vec{B}|$", where
$\vec{b}$ is the small spatially varying magnetic field on the
uniform background field $\vec{B}$.  Our choice for perturbative
kernel $d/h(r)$ with $d$ constant leads to a convenient separation of
orders.

The perturbative procedure is a functional separation method rooted
in the natural independence of the basis functions $\epsilon^n$
irrespective of their real amplitudes.  If the parameter $\epsilon$
is small almost everywhere, the series solutions will be convergent. 
There could be locations where the amplitudes of components seem to
be of differing orders as van Ballegooijen contends.  Where a
normally small expansion parameter $\epsilon$ actually becomes
relatively large series expansions may be divergent, but such
behavior characteristically defines the isolated singularities of
differential equations as discussed in theorems on the Frobinius
method for series solutions.

\section{Polytrope atmosphere}
\label{a:poly}

A polytrope atmosphere is defined by the pressure relation
$p(r)\propto \rho(r)^\Gamma$ for the polytrope adiabat $\Gamma$. Thus
the 1D polytrope scale height $h(r)= p(r)/(\rho(r) g(r))$ can be
written
 \begin{equation}
{h(r)\over h_s}=
\left({p(r)\over p(r_s)}\right)^{\Gamma-1\over\Gamma}
{g_s\over g(r)}=
\phi(r)^{\Gamma-1\over\Gamma}{g_s\over g(r)},
\label{e:poly}
\end{equation}
 making reference to a base height $r_s$ taken to be the stellar
surface, and introducing the scale-height function $\phi(r)$ for the
normalized gas pressure.

Substituting $h(r)$ into the hydrostatic Eq. (\ref{e:phi}) defines
the scale-height function for a polytrope
 \begin{equation}
\phi(r)^{-1/\Gamma}\ {\partial \phi(r)\over\partial r}=
-{g(r)\over g_s h_s},
\label{e:polydifa}
\end{equation}
 or
 \begin{equation}
{\partial \over\partial r} \phi(r)^{\Gamma-1\over\Gamma}=
-{\Gamma-1\over\Gamma}{g(r)\over g_s h_s},
\label{e:polydifb}
\end{equation}
 giving
 \begin{equation}
\phi(r)=\left[-{\Gamma-1\over\Gamma}\int{{g(r)\over g_s h_s}{\rm d}r}\right]^{\Gamma\over\Gamma-1},
\label{e:phipolya}
\end{equation}
 and with $g(r)=g_s(r_s/r)^2$, we obtain
 \begin{equation}
\phi(r)=\left[1-{\Gamma-1\over\Gamma}{r_s\over h_s}
\left(1-{r_s\over r}\right)\right]^{\Gamma\over\Gamma-1},
\label{e:phipolyb}
\end{equation}
 choosing the constant of integration so that $\phi(r_s)=1$.  The
part of the expression contained in brackets $[\ ]$ corresponds to
the radial temperature variation, the thermal lapse rate.  The
polytrope adiabat $\Gamma$ ranges from 1 to $5/3$ for an atmosphere
ranging from lossy and isothermal to adiabatically stratified.  For
$\Gamma=1$, the scale-height function $\phi(r)$ has a limiting pure
exponential decay form.  For $\Gamma>1$, the thermal lapse rate and
scale-height function have the unphysical feature that they become
negative at some height, showing that a polytrope of constant
$\Gamma$ can exist only over a limited range in $r$. Such behavior is
well known in stellar-interior polytrope models and leads to
convection zones of finite extent.

\section{Hydrodynamics around fixed fields}
\label{a:MHD}

If the magnetic field is dominant over the gas, and either the field
remains stationary, or the magnetic adjustment time is much shorter
than the hydrodynamic adjustment time, then the Lorentz term in the
MHD equation can be treated as constant in time and the hydrodynamics
separate from the magnetodynamics.

Fast magnetic adjustment is a remarkable feature of solar
reconnection phenomena. Sporadic crossings of coronal loops in
low-energy flare events (\citealt{LinJ+++1992}) suggest `X'-type
configurations, and the instability of the `X' configuration is well
known to form a new current sheet at near the Alfv{\'e}n velocity if
a nonzero resistivity can be affected in the region
(\citealt{Dungey1958, Syrovatskii1981}); such a resistivity is
predicted in some models (\citealt{PetschekHE+ThorneRM1967,
Priest1972}).  Taking a field strength of $|\vec{B}|\simeq$400 G as
representative of chromospheric magnetic fields near neutral lines in
solar active regions where flares commonly occur, we obtain an
Alfv{\'e}n speed $c_a= |\vec{B}|/ (4\pi\rho)^{1/2}\simeq 10^8$ cm
s$^{-1}$ (with $\rho= N \bar{m}=10^{-12}$ g cm$^{-3}$, using
$N=10^{12}$ cm$^{-3}$ and $\bar{m}=10^{-24}$ g). The Alfv{\'e}n speed
stays fairly constant into the corona giving a collapse time of
$\tau_B= h_B/ c_a\simeq$ 160 s for the propagation of the instability
through a magnetic scale height $h_B= 2h_{\rm cor}= r_s/(\ell+2)
\simeq 1.6\times 10^{10}$ cm, using a primary magnetic spherical
harmonic of $\ell=$2.33 in Eq. (\ref{e:hcor}).

The hydrodynamic response to already present or newly formed isolated
thin non-force-free flux tubes can be obtained by perturbative
analysis of the dynamical equations projected along the magnetic
field.  Taking the velocity to be aligned with the magnetic field and
of lesser order than the mean quantities adds a single
time-derivative term to the projected MHS Eq. (\ref{e:MHSg}), which
is written with the continuity equation
 \begin{eqnarray}
&&\hskip -.3cm\rho{\partial u_1\over\partial t}+{\partial p\over\partial z}
+\rho g (\vec{\nabla} z\cdot \vec{\nabla} r)=0,
\label{e:MHD}\\
&&\hskip -.3cm{\partial\rho\over\partial t}+{\partial\over\partial z}\left(\rho u_1\right)=0;
\label{e:cont}
\end{eqnarray}
 the equations are taken with the thermodynamic energy equation
$p\propto \rho^\gamma$; $z$ is the direction of the magnetic field;
$u_1$ denotes the velocity amplitude; $\gamma$ is the ratio of
specific heats, which ranges from 1 to 5/3 for lossy isothermal
refilling flows to adiabatic.  We now choose to use the density
$\rho$ as a problem variable, rather than the temperature or scale
height $h$.

Perturbing the quantities backward in time, $p=p_0-p_1$ and
$\rho=\rho_0-\rho_1$, from a final equilibrium state $(p_0,\rho_0)$
leaves the hydrostatic relation Eq. (\ref{e:phi}) with $p_0$ coupled
to the magnetic field through the cross-field Eq. (\ref{e:MHSgx}),
and three coupled perturbative equations for $(p_1,\rho_1,u_1)$
 \begin{eqnarray}
&&\hskip -.3cm\rho_0{\partial u_1\over\partial t}={\partial p_1\over\partial z}
+\rho_1 g \cos\theta,
\label{e:MHD1}\\
&&\hskip -.3cm{\partial\rho_1\over\partial t}={\partial\over\partial z}\left(\rho_0\ u_1\right),
\label{e:cont1}
\end{eqnarray}
 and $p_1=\gamma(p_0/\rho_0)\rho_1$, introducing $\theta$ as the
angle between the magnetic field direction $z$ and the radial $r$.
With small positive pressure and density deficits $p_1$ and $\rho_1$,
a small positive velocity $u_1$ is required in Eq. (\ref{e:MHD1})
corresponding to an upward flow opposite the normal gravitational
stratification.  Refilling changes the gas-pressure profile, in
effect heating the gas compressively as the dominant heating term in
the thermodynamic energy equation.

The unperturbed density $\rho_0=p_0/(gh)$ varies on a scale similar
to the gas pressure $p_0$, since the gravity $g$ and temperature
contained in $h$ vary on much larger scales; thus the derivative is
approximated $\partial \rho_0/\partial z \rightarrow \rho_0
\cos\theta/h$.  The perturbed quantities must vary on the scale of
the magnetic field $2h/\cos\theta$, giving $\partial
(p_1,u_1)/\partial z \rightarrow (p_1,u_1) \cos\theta/(2h)$. 
Substituting for the time derivative $\partial/\partial t \rightarrow
1/\tau_u$ lets us solve for the hydrodynamic timescale
 \begin{equation}
\tau_u= {1\over\cos\theta}
\left({4h\over(6+3\gamma)g}\right)^{1/2},
\label{e:tauu}
\end{equation}
 which is proportional to the gravitational free-fall time through
the scale height $h$.  For a radial field $\theta=0$, mixed lossy
conditions $\gamma=1.2$, and $h=r_s/(2\ell+4)$, we obtain
$\tau_u=$349 s for solar surface conditions with $\ell=2.33$, which
is just over twice the magnetic adjustment time $\tau_B$.  The
hydrodynamic adjustment time increases with more lossy conditions and
smaller $\gamma$, but overall depends only weakly on the
thermodynamics.


\end{document}